\documentclass[11pt]{article}

\usepackage{amssymb,amsmath,epsfig}
\include{epsf}

\usepackage[hypertex]{hyperref}

\newcommand{\be}{\begin{equation}}
\newcommand{\ee}{\end{equation}}
\def\beqa{\begin{eqnarray}}
\def\eeqa{\end{eqnarray}}
\def\nn{\nonumber}
\newcommand{\Rl}{\mathbb{R}^3_\lambda}
\newcommand{\Rt}{\mathbb{R}^4_\theta}
\newcommand{\R}{\mathbb{R}}
\newcommand{\C}{\mathbb{C}}
\newcommand{\N}{\mathbb{N}}

\newcommand{\eqn}[1]{(\ref{#1})}
\newcommand{\del}{\partial}
\newcommand{\Tr}[1]{\:{\rm Tr}\,#1}

\renewenvironment{thebibliography}[1]
         {\section*{References}\frenchspacing\small
          \begin{list}{[\arabic{enumi}]}
         {\usecounter{enumi}\parsep=2pt\topsep 0pt
         \settowidth{\labelwidth}{[#1]}
         \leftmargin=\labelwidth\advance\leftmargin\labelsep
         \rightmargin=0pt\itemsep=1pt\sloppy}}{\end{list}}

 \numberwithin{equation}{section}

\title{Noncommutative field theory on $\R^3_\lambda$}
\author{Patrizia Vitale$^{a,b}$}

\begin{document}
\maketitle

\vspace*{-1cm}

\begin{center}

\textit{$^a$Dipartimento di Fisica
Universit\`a di Napoli Federico II}  \\
\textit{$^b$INFN, Sezione di Napoli, Via Cintia 80126 Napoli, Italy}\\
 e-mail:
\texttt{patrizia.vitale@na.infn.it}\\[1ex]

\end{center}

\begin{abstract}
We consider the noncommutative space $\Rl$, a deformation of the algebra of functions on $\R^3$ which yields a foliation of $\mathbb{R}^3$ into fuzzy spheres. We first review the construction of  a natural matrix basis adapted to $\Rl$.  We thus consider the problem of defining a new Laplacian operator for the deformed algebra. We propose an operator which is not of Jacobi type.   The implication for field theory of the new Laplacian is briefly discussed.
\bigskip

\end{abstract}

\pagebreak

\section{Introduction.}
There are many well known arguments which support the relevance of noncommutative geometry in physics (for a quick recollection, see for example \cite{wiki}). They mainly deal with the  nature of space-time at Plank scale, where the quantization of the gravitational field should become non-negligible. A very simple and beautiful argument may be found for example in \cite{Doplich1} (also see \cite{Bronstein}),  which brings into question  the smooth structure of space-time at very small scales,  when both quantum mechanics and general relativity are taken into account.
Space-time non commutativity  emerges from very different approaches to the quantization of the gravitational field, such as Loop Quantum gravity, where a minimal value is found in the spectrum of the area operator \cite{LQG}, string theory, where the discovery of space-time noncommutativity in the presence of a background field gave a great boost to the research in noncommutative geometry \cite{SW} and, recently, double string theory (see \cite{lust} for a review), with space-time noncommutativity emerging from a manifestly T-dual invariance of the classical action. Moreover, it can be advocated as a regularization tool in QFT, as it was pointed out in the  very beginning of quantum field theory by Heisenberg and Snyder.

The renormalization study of  NCQFT (Noncommutative Quantum Field Theory)  is in general difficult and is often complicated by the Ultraviolet/Infrared (UV/IR) mixing \cite{Minwalla}. The phenomenon persists in Moyal-noncommutative gauge models and represents one of the main open problems of Moyal-based field theory.
A first solution for scalar field theory was  proposed in 2003 \cite{Grosse:2003aj}. It  amounts to modify the initial action with a harmonic oscillator term leading to a fully renormalisable NCQFT.
This is the so called Grosse-Wulkenhaar model.

There is a large production on NCQFT with Moyal non commutativity (for a review see for example \cite{douglas, szabo} and references therein). In the search of renormalizable NCQFT, in this paper we pursue a different approach, which, instead of modifying the action within Moyal noncommutativity, aims at investigating different kinds of noncommutativity. It  is based on a previous article of the author in collaboration with J.-C. Wallet \cite{VW12} where a family of scalar quantum field theories on the noncommutative space $\R^3_\lambda$  is considered. These models have been recently  extended to the gauge setting in \cite{GVW14}. The noncommutative algebra, known as $\R^3_\lambda$ was first introduced in \cite{Hammaa} (also see \cite{selene}).
A useful starting point consists in describing  the algebra of functions on $\R^3$ as a sub-algebra of functions on $\R^4$. The map $\R^4\rightarrow \R^3$ (with zero points removed) is a  fibration, with compact fiber $U(1)$.  It is known in the commutative  setting as   the Kunstaanheimo-Stiefel map \cite{KS}. It can be dualized on using the pull-back map and a differential and integral calculus can be derived \cite{DMV05}.  We shall follow this approach in the present paper, to perform a critical review of \cite{VW12} and address some open problems which emerged in \cite{VW12},\cite{GVW14}, such as the issue of the integration measure and the algebra of derivations. An open  problem in \cite{VW12} was the fact that the proposed Laplacian, based on the inner derivations of the algebra, didn't describe the radial dynamics. In other terms, being constructed with derivations which in the commutative limit correspond to  vector fields tangent to the spheres of the foliation, it couldn't describe in a satisfactory manner a three-dimensional dynamics. As we shall see in some detail, the problem is due to the fact that the radial derivation is not a $\star$-derivation, unless we enlarge the algebra $\R^3_\lambda$. We briefly discuss such a possibility, but we stick to the idea, already contained in \cite{VW12} but not fully exploited, that the radial dynamics may be implemented by means of the multiplication by the radial coordinate $x_0$, without modifying the algebra. We thus propose a new Laplacian, which implements radial dynamics. The commutative limit of our model is however still unlcear to us. 

The paper is organized as follows. In section \ref{sect2} we summarize the general properties of the noncommutative $\Rl$ that will be used in this paper together with some features related to the Wick-Voros product. In  section \ref{matrixbasesection} we review the  natural matrix basis adapted to $\Rl$, we define an integration measure and  the derivations of the algebra.  In  section \ref{theactions} we construct a new Laplacian. We thus discuss the implications of the new Laplacian for a family of real-valued scalar actions on $\Rl$ with quartic polynomial interactions. We conlude with a list of open problems. \par

\section{The noncommutative algebra $\Rl$}\label{sect2}
The noncommutative space $\Rl$ has been first introduced in \cite{Hammaa}. A generalization has been studied in \cite{selene}.  It is  a subalgebra of $\Rt$, the noncommutative algebra of functions on $\R^4\simeq \C^2$ endowed with the Wick-Voros product \cite{Wick-Voros}
\be
\phi\star \psi\, (z_a,\bar z_a)= \phi(z,\bar z) \exp(\theta \overleftarrow\del_{z_a}\overrightarrow\del_{\bar z_a}) \psi(z,\bar z), \,\,\,\, a=1,2 \label{Wick-Vorosprod}
\ee
  For coordinate functions we thus  have the $\star$-commutator
$
[z_a,\bar z_b]_\star= \theta \delta_{ab}\,\;
$
with $\theta$ a constant, real parameter.
The crucial step  to obtain star products on $\mathcal{F}(\mathbb{R}^3)$, hence to deform $\mathcal{F}(\mathbb{R}^3)$ into a noncommutative algebra, is to identify
$\mathcal{F}(\mathbb{R}^3)$ with a subalgebra of $\mathcal{F}(\mathbb{R}^4)$. This is achieved on identifying the coordinates of $\R^3$ as quadratic functions in $\bar z_a, z_a$
\be
x_\mu=\frac{1}{2} \bar z^a \sigma_\mu^{ab} z^b, \;\;\; \mu=0, ..,3 \label{xmu}
\ee
with      $\sigma_i$
 the Pauli matrices, while $\sigma_0$ is the identity. The subalgebra polynomially generated by the coordinate functions $x_\mu$ and properly completed, is closed with respect to the Wick-Voros star product. We have indeed
 \be
 [x_i,x_j]_\star= i \lambda \epsilon_{ij}^k x_k. \label{starcom}
 \ee
In order to have the right dimensions for the coordinates $x_i$, we choose in this paper $z_a$ to have length dimension of $1/2$, so that we can choose $\lambda=\theta $, of length dimension $1$, and there is no need to introduce other dimensional constants, differently from \cite{VW12}.
 Notice that
\be
\sum_i x_i^2 = x_0^2 \label{x02}
\ee
and $x_0$ star-commutes with $x_i$. Thus  we can alternatively define $\R^3_\lambda$ as the star-commutant of $x_0$.
It is easily verified that the induced  $\star$-product for $\R^3_\lambda$ reads then
\be
\phi\star \psi \,(x)= \exp\left[\frac{\lambda}{2}\left(\delta_{ij} x_0+ i \epsilon_{ij}^k x_k \right)\frac{\del}{\del u_i}\frac{\del}{\del v_j}\right] \phi(u) \psi(v)|_{u=v=x} \label{starsu2}
\ee
which gives, for coordinate functions
\beqa
x_i\star x_j &=& x_i x_j+ \frac{\lambda}{2} \left(x_0 \delta_{ij} + i \epsilon_{ij}^k x_k\right) \label{xistarxj}\\
x_0\star x_i &=& x_i\star x_0 = x_0 x_i + \frac{\lambda}{2} x_i\\
x_0\star x_0&=& = x_0(x_0+\frac{\lambda}{2}) = \sum_i x_i\star x_i - \lambda x_0  \label{x0*2}
 \eeqa
 in agreement with \eqn{starcom}.
  Here Eq.  \eqn{x02} has been used.  The product is associative, since it is nothing but the Wick-Voros product expressed in different variables.
 The commutative limit is achieved with $\lambda\rightarrow 0$.

Let us point out that the identification of the algebra $\R^3_\lambda$
 as a subalgebra of $\R^4_\theta$ has a geometric counterpart in the commutative setting, where the Kustaanheimo-Stiefel (KS) map \cite{KS} can be used. The main idea there, is the observation that $\R^3-\{0\}$ and $\R^4-\{0\}$ may be given the structure of trivial bundles over spheres, being $\R^3-\{0\}\simeq S^2\times\R^+$ and $\R^4-\{0\}\simeq S^2\times \R^+$. Then one may use the well known Hopf fibration $\pi_H : S^3\rightarrow S^2$, with the identification of $S^3$ with $SU(2)$,
 \be
 \pi_H: s\in SU(2)\rightarrow \vec x \in S^2, \;\; : s\sigma_3 s^{-1}= x^i \sigma_i
 \ee
 where $s= y_0\sigma_0+ i y_i \sigma_i$ and $y_\mu$ are real coordinates on $\R^4$ such that $y_\mu y^\mu=1$. Now one may extend (not uniquely)  the Hopf  map to $\R^4-\{0\}\rightarrow \R^3-\{0\}$, relaxing the radius  constraint so that $y_\mu y^\mu=R^2$, with $R\in \R^+$. On introducing $g= R s$ we define
 \be
 \pi_{KS}: g \in\R^4-\{0\}\rightarrow \vec x\in \R^3-\{0\}, \;\;\; x^k\sigma_k=g \sigma_3 g^\dag=  R^2 s\sigma_3 s^{-1}. \label{KS}
 \ee
 One can easily verify that this map gives back relations \eqn{xmu} up to a factor of $2$, with $z_1= \frac{1}{\sqrt{2}} (y_0+i y_3)$, $z_2=\frac{1}{\sqrt{2}}(y_1+i y_2)$ and the identification
 \be x_0=\frac{ R^2}{4}. \label{rR}
 \ee
 The KS fibration may be used to define the derivations for the algebra of functions $\mathcal{F}(\R^3-\{0\})$ as projections of the derivations of $\mathcal{F}(\R^4-\{0\})$ \cite{DMV05}. We shall see that this procedure can be generalized to  the noncommutative setting. Moreover, with the introduction of the matrix basis, the restriction to $\R^3-\{0\}$ may be removed, since we shall see that our matrix basis is well defined in $0\in \R^3$ as well.

\section{\texorpdfstring{The matrix basis}{Matrix basis }}\label{matrixbasesection}

In \cite{VW12} a
matrix basis for the algebra $\R^3_\lambda$ has been introduced,  as a reduction of the matrix basis of the Wick-Voros algebra $\R^4_\theta$ \cite{discofuzzy, V-T}.
The latter is similar to the one introduced for the Moyal $\R^4_\theta$ \cite{pepejoe}. In 2-d it is based on the expansion
\be
\phi(\bar z,z)=\sum_ {pq}\tilde\phi_{pq} \bar z^p z^q\, ,\;\; p,q\in \mathbb{N} \;\;\; \tilde\phi_{pq}\in \C
\ee
to which we associate the  normal ordered operator
\be
\hat \phi =\sum_ {pq}\tilde\phi_{pq}  a^{\dag p} a^q ,\;\;\,\; \tilde\phi_{pq}\in\C \label{normord}
\ee
with $a, a^\dag$  space-time  creation and annihilation operators. The inverse map is simply
\be
\phi(z,\bar z)= \langle z|\hat \phi |z\rangle \label{wign}
\ee
with $|z\rangle$ the coherent state associated to the operator $a$. Eq. \eqn{wign}  defines the Wick-Voros star product as
\be
\phi\star\psi (z,\bar z)=  \langle z| \hat\phi\hat\psi|z\rangle \label{vorosprod}
\ee
which is associative by construction. It can be given an integral form in terms of an integral kernel (see \cite{discofuzzy} for details). It is however mostly known in its asymptotic form, already presented in Eq. \eqn{Wick-Vorosprod}.

Thus we generalize to  four dimensions, with  $a_a, a_a^\dag$, $a=1,2$ and re-express the operator $\hat \phi$ in  the harmonic oscillator basis, $|N\rangle:=|n_1,n_2\rangle$
\beqa
a_1|n_1,n_2\rangle&=&{\sqrt{\theta}}{\sqrt{n_1}}|n_1-1,n_2\rangle,\ {{a_1^\dag}}|n\rangle={\sqrt{\theta}}{\sqrt{n_1+1}}|n_1+1,n_2\rangle,\\
 a_2|n_1,n_2\rangle&=&{\sqrt{\theta}}{\sqrt{n_2}}|n_1,n_2-1\rangle,\ {{a_2^\dag}}|n\rangle={\sqrt{\theta}}{\sqrt{n_2+1}}|n_1,n_2+1\rangle
\eeqa
so to obtain
\be
\hat\phi=\sum_{P,Q\in\mathbb{N}^2}\phi_{PQ} {|P\rangle \langle Q| }\, , \; \;\;\;\;  |P\rangle:=|p_1,p_2\rangle =\frac{{a}_1^{\dag p_1}{a}_2^{\dag p_2}} {[P!\theta^{|P|}]^{1/2} }|0\rangle,\;\; \;  \forall P=(p_1,p_2)\in\mathbb{N}^2 \label{changeofb}
\ee
with $\phi_{PQ}\in \C$ 
and $|P|=p_1+p_2$, $P!=p_1! p_2!$. 
Thus the matrix basis in $\R^4_\theta$ is obtained through Eq. \eqn{wign}
\be
{f_{PQ}(z,\bar z)}= \langle z_1,z_2|{\hat f_{PQ}}|z_1,z_2\rangle=
\frac{e^{-\frac{ \bar z_1 z_1+\bar z_2 z_2}{\theta}}}{\sqrt{P!Q!\theta^{|P+Q|}}} \bar z_1^{p_1}\bar z_2^{p_2}  z_1^{q_1} z_2^{q_2}
\ee
with ${\hat f_{PQ}:=|P\rangle \langle Q| } $. The following fusion rule and orthogonality relations hold
\be
f_{MN}\star f_{PQ}(z,\bar z)= \delta_{NP} f_{MQ}(z,\bar z)\, , \;\;\;\;
\int d^2 z_1 d^2 z_2\, f_{PQ}(z,\bar z)= (2\pi \theta)^2 \delta_{PQ}
\ee
so that the star product in $\R^4_\theta$  becomes a matrix product
\be
 \phi\star\psi (z, \bar z)= \sum \phi_{MN}\psi_{PQ} f_{MN}\star f_{PQ}= \sum  \phi_{MP}\psi_{PQ} f_{MQ}
\ee
and the integral  becomes a trace
 \be
 \int \phi\star \psi \star ...=  (2\pi \theta)^2 \Tr \Phi \Psi ... \label{trace}
 \ee
$\Phi=(\phi_{MN}), \Psi= (\psi_{MN})$, are infinite matrices.  Thus the trace \eqn{trace} needs to be regularized when used in the context of field theory.

The  matrix basis of  $\R^3_\lambda$
is obtained as a reduction from the previous one, using the association of the coordinate functions $x_\mu$ with the $SU(2)$ generators and their realization  in terms of creation and annihilation operators. This is nothing but the quantum version of the KS map.
Indeed, given the number operators $\hat N_1=a^\dag_1 a_1, \; \hat N_2=a^\dag_2 a_2$ with eigenvalues $n_1, n_2$ we have the well known relation
\be n_1+n_2= 2j \;\;\; n_1-n_2= 2m
\ee
with $j(j+1)$ and $m$ eigenvalues of $\sum_i \hat X_i \hat X_i$ and $\hat X_3$ respectively.
We have then
\be
|n_1,n_2\rangle\longrightarrow |j+m, j-m\rangle
\;\; {\rm and}\;\;\;
{ \hat f_{NP}}= |n_1,n_2\rangle\langle p_1, p_2| \longrightarrow  |j+m, j-m\rangle \langle \tilde\jmath+\tilde m, \tilde \jmath-\tilde m | \equiv{\hat v^{j\tilde\jmath}_{m\tilde m}}.
\ee
This implies
$
{ f_{NP}(\bar z, z)}\longrightarrow  {v^{j\tilde\jmath}_{m\tilde m}(\bar z, z)}.
$ 
 For this to be a basis in $\R^3_\lambda$ we finally impose it to $\star$-commute with $x_0$
\be
 x_0 \star v^{j\tilde\jmath}_{m\tilde m} (z,\bar z)-v^{j\tilde\jmath}_{m\tilde m}\star
x_0 (z,\bar z)=\lambda(j-\tilde\jmath) v^{j\tilde\jmath}_{m\tilde m}
 \ee
 This fixes     ${ j=\tilde\jmath}$ and we pose $v^j_{m\tilde m}:= v^{jj}_{m\tilde m}$.
 We have then
\be\phi(x)=\sum_{j\in \frac{\N}{2}}\sum_{m,\tilde m=-j}^j \phi^j_{m\tilde m} v^j_{m\tilde m}(x) \label{funexp}
\ee
with
\be
v^j_{m\tilde m}(x)= \frac{ e^{-2\frac{ x_0}{\lambda}}}{ \lambda^{2j}}  \frac{(x_0+x_3)^{j+m}
(x_0-x_3)^{j-\tilde m}\; (x_1-i x_2)^{\tilde m -m}
}{\sqrt{(j+m)!(j-m)! (j+\tilde m)!(j-\tilde m)! }}.
\ee
Notice however that this expression is not unique, because of the relations between the coordinates $x_\mu$.

Let us notice that the functions expansion \eqn{funexp} is well behaved in $x=0$, it being $\lim_{x\rightarrow 0}\phi(x) = \phi^0_{00}$ .  Thus, on redefining $\tilde x= x/\lambda$ , $v^{j}_{m\tilde m}(x)$ provides a basis for the commutative and noncommutative algebras of functions on the whole $\R^3$ , analogously to the Moyal matrix basis \cite{pepejoe},  under usual regularity assumptions for  the sequence of the coefficients $\{\phi^j_{m\tilde m}\}$.
The star product \eqn{starsu2} applied to the basis elements acquires the simple form
\be
v^j_{m\tilde m}\star v^{\tilde\jmath}_{n \tilde n}(x)=\delta^{j \tilde\jmath}\delta_{\tilde m
n}v^j_{m \tilde n}(x) \label{matrixprod}
\ee
Then, the star product in $\mathbb{R}_\lambda^3$ becomes a block-diagonal infinite-matrix product
\be
{\phi\star \psi}(x)=\sum_{j,m_1, \tilde m_2} \phi^{j}_{m_1\tilde m_1} \psi^{j}_{m_2\tilde m_2}
v^{j}_{m_1\tilde m_1} \star v^{j}_{m_2\tilde m_2} \; = {\sum_{j,m_1, \tilde m_2} (\Phi^j\cdot \Psi^j)_{m_1 \tilde m_2} v^j_{m_1 \tilde m_2}} \label{starprodtr}
\ee
 the infinite matrix $\Phi$ gets rearranged into a block-diagonal form, each block being the $(2j+1)\times (2j+1)$ matrix   {$\Phi^j=\{\phi^j_{mn}\}, \, -j\le m,n\le j$}.
\subsubsection{The matrix representation of coordinates}
The coordinate functions may be expressed in the matrix basis and their action on the basis is easily computed.
On using the expression of the generators in terms of $\bar z_a, z_a$, \eqn{xmu}, we find
\be\label{coordinates}
\begin{array}{lll}
x_+ = \lambda \sum_{j,m}\sqrt{(j+m)(j-m+1)} v^j_{m\,m-1} & &
x_- =  \lambda \sum_{j,m}\sqrt{(j-m)(j+m+1)} v^j_{m\,m+1}\\
x_3 =  \lambda \sum_{j,m} m v^j_{m\,m}& &
x_0 = \lambda \sum_{j,m} j v^j_{m\,m}
\end{array}
 \ee
were we have introduced
$
x_\pm= x_1\pm i x_2.
$
Thus we compute
\be \label{x0v}
\begin{array}{lll}
x_+\star v^j_{m\tilde m}= \lambda \sqrt{(j+m+1)(j-m)} v^j_{m+1 \, \tilde m}  & & v^j_{m\tilde m}\star x_+= \lambda\sqrt{(j-\tilde m+1)(j+\tilde m)} v^j_{m\, \tilde m -1} \\
x_-\star v^j_{m\tilde m}= \lambda \sqrt{(j-m+1)(j+m)} v^j_{m-1\,\tilde m}  &&v^j_{m\tilde m}\star x_-= \lambda\sqrt{(j+\tilde m+1)(j-\tilde m)} v^j_{m \,\tilde m +1} \\
x_3\star v^j_{m\tilde m}= \lambda\, m \, v^j_{m\tilde m}&&  v^j_{m\tilde m} \star x_3= \lambda \,\tilde m \, v^j_{m\tilde m}\\
x_0\star v^j_{m\tilde m}= \lambda\, j \, v^j_{m\tilde m}&&  v^j_{m\tilde m} \star x_0= \lambda\, j \,v^j_{m\tilde m}
\end{array}
 \ee
Therefore, the basis we have chosen diagonalizes the two coordinates $x_0$ (the radius) and $x_3$ with eigenvalues respectively $\lambda j$ and $\lambda m$ ($\lambda\tilde m$ when acting on the right).
 \subsection{Integration}
 The definition of the integral in the algebra $\R^3_\lambda$ is based on \cite{DMV05}, where the  KS map is used. They  observe  that $\pi_{KS}$, defined in \eqn{KS}, defines a principal fibration $\R^4-\{0\}\rightarrow \R^3-\{0\}$ with structure group $U(1)$. The fibre is therefore compact. Moreover $\mathcal{F}(\R^3-\{0\})$ can be embedded into $\mathcal{F}(\R^4-\{0\})$ by
 \be
\pi^*_{KS}: f \in \mathcal{F}(\R^3-\{0\})\rightarrow f\circ \pi_{KS}\in \mathcal{F}(\R^4-\{0\})
\ee
with  $\pi_{KS}^\star$ the pull-back map.
This realizes $\mathcal{F}(\R^3-\{0\})$ as the subalgebra of $\mathcal{F}(\R^4-\{0\})$ of functions which are constant along the fibers.
The vector field which generates the fiber $U(1)$
\be
Y_0= y^0 \frac{\del}{ \del y^3} - y^3 \frac{\del}{ \del y^0 } +y^1 \frac{\del}{ \del y^2} -y^2 \frac{\del}{ \del y^1} \label{fibergen}
\ee
defines indeed $\mathcal{F}(\R^3-\{0\})$ as its kernel (it corresponds in the noncommutative case to the inner derivation $[x_0, \cdot]_\star$).
It is shown that, given the ordinary volume forms on $\R^3$ and $\R^4$, $\mu_3= dx^1\wedge dx^2\wedge dx^3$ and $\mu_4= dy^0\wedge dy^1\wedge dy^2\wedge dy^3 $, we have
\be
\pi^\star_{KS} (\mu_3 ) \wedge \alpha_0\propto  R^2 \mu_4
\ee
with $\alpha_0$ the dual form of the vector field $Y_0$ and volume form on the fiber. With our conventions the proportionality factor is $1/4$.
Observing that functions on $\R^3$ are constant along the fiber $U(1)$, we can factorize the integral along the fiber which just gives a factor of $2\pi$, so that
 we have
 \be
\int \mu_3 \;f =\frac{1}{2\pi}\int \mu_4\; \pi^*_{KS}(x_0) \;  \pi^*_{KS}(f) \label{intdef}
\ee
where \eqn{rR} has been used.

Therefore, we may generalize to  the noncommutative case, assuming    \eqn{intdef}  as a definition,
\be
\int_{\R^3_\lambda} f:=\frac{1}{2\pi}\int_{\R^{4}_\theta } \pi^*_{KS}(x_0) \;  \bullet  \pi^*_{KS}(f)  \label{intnoncom}
\ee
compatible with the commutative limit. The symbol $\bullet$ indicates the two possible choices that we have in generalizing \eqn{intdef}, which correspond to  star-multiply  the weight coming from the integration measure, $\pi^*_{KS}(x_0)$,  with the integrand  $ \pi^*_{KS}(f)$ or to use the commutative product. Since $x_0$ star-commutes with all elements of the algebra, the two definitions only differ by a constant shift, as we shall see in a moment. Let us point out that \eqn{intnoncom} is different from the one in   \cite{VW12}, where the factor of $\pi^*_{KS}(x_0)$ is absent.
Both definitions are legitimate, but the one proposed here has the advantage of reproducing the usual integral on $\R^3$ once the commutative limit is performed.
Eq. \eqn{intnoncom} implies for the basis functions
\be
\int v^j_{m\tilde m} (x)= \left\{
\begin{array}{ll}
8\pi \lambda^3 j\; \delta_{m \tilde m}   &a)\\
8\pi\lambda^3 (j+1)\; \delta_{m \tilde m}& b)
\end{array}
\right.
\ee
where the result a) corresponds to  the choice to star-multiply the weight-function $x_0$ with the integrand whereas the result b) corresponds to choosing the point-wise multiplication. As announced, it amounts to a constant shift. We shall choose the second option in the paper.
We thus have
\be
\int v^j_{m\tilde m}  \star v^j_{n\tilde n} (x)   =
8\pi\lambda^3(j+1)\delta_{\tilde m n }\delta_{m \tilde n}
\ee
Thanks to these results we obtain, for the integral of the star product \eqn{starprodtr}
\be
\int_{\mathbb{R}^3_\lambda} \phi\star \psi = 8\pi\lambda^3\sum_j (j+1) \Tr_j (\Phi^j\cdot \Psi^j)
\ee
with $\Tr_j$ the trace in the $(2j+1)\times(2j+1)$  subspace.
Notice that, on performing the sum up to $j=N/2$, with $\Phi^j=\Psi^j=  \bf{1}_j$ we obtain the result $8\pi\lambda^3\sum_{n=0}^N(n/2+1)(n+1)\simeq\frac{4}{3}\pi\lambda^3 N^3$, which reproduces correctly  the volume of a sphere of radius $\lambda N$.
\subsection{Derivations}
In order to introduce a dynamics described by an action functional we need derivations. In the commutative case they are obtained by projecting the derivations of the algebra $\mathcal{F}(\R^4)$ through the KS map \eqn{KS}. It may be seen \cite{DMV05} that projectable vector fields are defined by the condition [$D_i, Y_0]=0$, with $Y_0$ given in \eqn{fibergen} . They correspond to the three rotations generators and the dilation
\be
\pi_{KS*} (Y_i)= X_i=  \epsilon_{ijk}x_j \frac{\del}{\del x_k}\, , \;\;\;\;
\pi_{KS*} (D)= x_i\frac{\del}{\del x_i} \label{dilation}
\ee
with $Y_i =y_0 \frac{\del}{\del y_i}  -y_i \frac{\del}{\del y_0} -\epsilon_{ijk}y_j  \frac{\del}{\del y_k}$ and $D= y_\mu\frac{\del}{\del y_\mu} $. The three rotations are not independent since $x_i \cdot X_i =0$. When passing to the noncommutative case the three rotations  are still derivations of the algebra $\R^3_\lambda$ and may be given the form of inner derivations, but the dilation, in the form of \eqn{dilation}, is not anymore a derivation.
On using the star product \eqn{starsu2} we have for rotations
\be
X_i (\phi)= -\frac{i}{\lambda}[x_i,\phi]_\star\, ,  \,\,\; i=1,..,3
\ee
which obviously satisfies the Leibnitz rule. Moreover they become independent (even though $x_i \star X_i (\phi)+   X_i (\phi)\star x_i=0$, derivations are not a module over the algebra in the NC case, they are only a left module over the center of the algebra). As for the dilation, it is easy to check that it does not satisfy the Leibnitz rule (for example, on applying it to the star product of coordinates \eqn{xistarxj}-\eqn{x0*2}).

However, let us notice that there is a way to implement the dilation as a derivation of the star product \eqn{starsu2}. This amounts to enlarge the algebra $\R^3_\lambda$ to include the noncommutativity parameter $\lambda$ (see \cite{GLRV06} where the construction is performed for the Moyal algebra $\R^4_\theta$). It may be checked that, in such a case, the vector field
\be
D_\lambda= x_i\frac{\del}{\del x_i} + \lambda \frac{\del}{\del \lambda}\equiv x_0\frac{\del}{\del x_0} + \lambda \frac{\del}{\del \lambda}
\ee
is an outer derivation of the enlarged algebra.
We have indeed
\be
D_\lambda v^{j}_{m\tilde m}=0
\ee
it being $v^{j}_{m\tilde m}=v^{j}_{m\tilde m}(x/\lambda)$. This implies
\be
D_\lambda \phi= \sum_{j m\tilde m} \lambda\frac{\del}{\del \lambda}\phi^j_{m\tilde m} v^{j}_{m\tilde m}
\ee
with the coefficients $\phi^j_{m\tilde m}$ depending on $\lambda$ (see for example the coefficients  of the coordinate functions in the matrix basis expansion \eqn{coordinates}).
It is then easy to verify that the Leibnitz rule is satisfied
\be
D_\lambda (\phi\star\psi)= \sum \lambda\frac{\del}{\del \lambda}(\phi^j_{mn}\psi^j_{n\tilde n})v^{j}_{m\tilde n}= D_\lambda \phi\star\psi+ \psi\star D_\lambda \psi.
\ee

We shall investigate this new derivation and the possibility of building a new Laplacian on the enlarged algebra elsewhere. It will probably require a modification of the integration measure.  In the present paper  we shall stick to the  three independent derivations $X_i$, which are the only derivations of the algebra $\R^3_\lambda$ within the present definition.

\section{The Laplacian}\label{theactions}
It was already noticed in \cite{VW12} that  the dilation, which is needed to describe radial dynamics, is contained in the star multiplication by the coordinate $x_0$
\be
x_0\star \phi=  x_0 \phi + \frac{\lambda}{2} x_i \del_i \phi \label{radialgenerat}
\ee
It was thus suggested that the Laplacian operator should contain a term of the kind $x_0\star x_0$ so to have a term of second order in the radial derivative. There were however two problems. The first is that the relation to the ordinary commutative  Laplacian was unclear, the second, is that the new term, $x_0\star x_0$ is  diagonal in the matrix basis $v^j_{m\tilde m}$, with eigenvalue $\lambda^2 j^2$. This means that the wanted radial dynamics (which should instead affect the radius, hence it should change $j$) is not obtained.

Here we tackle the second problem, whereas the issue of the commutative limit is still unclear.  We argue  that, in order to produce an action on the radius eigenvalue $j$,  the correct term to consider is indeed the one {\it without} star product $
x_0^2 \phi$. Let us compute its action on the matrix basis. We find
\be
x_0^2 v^j_{m\tilde m}= {\lambda^2}  \left( M^j_{m \tilde m} v^{j+1}_{m+1\,\tilde m+1}+ N ^j_{m \tilde m}  v^{j+1}_{m\,\tilde m}+ Z^j_{m \tilde m}  v^{j+1}_{m-1\,\tilde m-1}\right)
\ee
with
\beqa
M^j_{m \tilde m}&=& \frac{1}{4} \sqrt{(j+m+2)(j+m+1)(j+\tilde m+2 )(j+\tilde m+1) } \nn\\
N^j_{m \tilde m}&=& \frac{1}{2} \sqrt{(j+m+1)(j-m+1)(j+\tilde m+1 )(j-\tilde m+1) } \nn\\
Z^j_{m \tilde m}&=& \frac{1}{4} \sqrt{(j-m+2)(j-m+1)(j-\tilde m+2 )(j-\tilde m+1) }. \label{MNZ}
\eeqa
As we can see, it increases   the index j of the matrix basis.
The proposed  Laplacian acting on a scalar field $\phi$  is therefore
\be
\Delta \phi = \frac{1}{x_0\star x_0}\star\left ( \alpha \sum_i X_i^2\phi + {\beta}\; \frac{x_0^2}{\lambda^2} \phi\right) \label{lapl}
\ee
with $\alpha, \beta$ real parameters.
The star-inverse of $x_0$ is defined by its action on the matrix basis, as in Eqs. \eqn{x0v}. As we shall see, it essentially amounts to divide by $j$.
In the following we assume the scalar fields $\phi \in R^3_\lambda$ to be real. 
\subsection{The scalar action}\label{subsect4.1}
Let us consider a scalar model with quartic interaction
\be
S[\phi]=\int  \phi\star(\Delta+\mu^2) \phi + \frac{g}{4!}\phi\star\phi\star\phi\star\phi \label{action}
\ee
The fields have length dimension equal to -1/2. This action was already considered in \cite{VW12}, though with a different Laplacian. 
On using the expansion of the fields in the matrix basis \eqn{funexp} and the multiplication rule for the basis elements \eqn{matrixprod}
we obtain the
action  in \eqn{action} as a matrix model action
\beqa
S[\phi]&=&{8\pi \lambda}\Bigl\{\sum \phi^{j_1}_{m_1\tilde m_1}\bigl(\Delta+\mu^2\mathbf{1}\bigr)^{j_1 j_2}_{m_1 \tilde m_1; m_2\tilde m_2} \phi^{j_2}_{m_2\tilde m_2} \Bigr.\nn\\
&+&\Bigr. {{g}\over{4!}}\sum \phi^{j_1}_{mn} \phi^{j_2}_{np} \phi^{j_3}_{pq} \phi^{j_4}_{qm} (j_1+1)\delta_{j_1 j_2} \delta_{j_2 j_3} \delta_{j_3 j_4}\Bigr\}
\label{action1}
\eeqa
where sums are understood over all the indices.  As already noticed in \cite{VW12} the interaction term, being polynomial in the fields, is diagonal in the matrix basis. It amounts to take the trace of the product $\Phi^j \Phi^j \Phi^j \Phi^j$ in the $(2j+1)\times (2j+1)$ subspaces and sum over j with a weight (2j+1), which comes form the integration measure. This weight was absent in \cite{VW12} because the integration measure was different.

The kinetic operator is not diagonal. It may be verified  to be
\beqa
&&(\Delta+\mu^2\mathbf{1})^{j_1 j_2}_{m_1\tilde m_1;m_2\tilde m_2}= \int_{\R^3_\lambda} v^{j_1}_{m_1 \tilde m_1}\star( \Delta +\mu^2\mathbf{1}) v^{j_2}_{m_2\tilde m_2}=\frac{j_1+1}{ j_1^2 }\Bigl\{ \alpha\delta^{j_1 j_2}\Bigl( \Bigr.\Bigr.
\nn\\
&& \;\;\;
\delta_{\tilde m_1 m_2}\delta_{m_1 \tilde m_2} D^{j_2}_{m_2\tilde m_2}-\delta_{\tilde m_1, m_2+1}\delta_{m_1, \tilde m_2+1}B^{j_2}_{m_2,\tilde m_2} - \delta_{\tilde m_1, m_2-1}\delta_{m_1, \tilde m_2-1} H^{j_2}_{m_2,\tilde m_2}\,\Bigr) + \beta \delta^{j_1j_2+1} \nn\\
&& \;\;\Bigl. \Bigl(
\delta_{\tilde m_1 m_2+1}\delta_{m_1 \tilde m_2+1} M^{j_2}_{m_2\tilde m_2}+\delta_{\tilde m_1, m_2}\delta_{m_1, \tilde m_2}N^{j_2}_{m_2,\tilde m_2} + \delta_{\tilde m_1, m_2-1}\delta_{m_1, \tilde m_2-1} Z^{j_2}_{m_2,\tilde m_2}\,\Bigr)\Bigr\}
\label{kineticterm}
\eeqa
with $M^{j_2}_{m_2\tilde m_2}, N^{j_2}_{m_2\tilde m_2},Z^{j_2}_{m_2\tilde m_2}$ given in Eqs. \eqn{MNZ} and
\beqa
D^j_{m_2\tilde m_2}&=&[  2(j^2 +j- m_2 \tilde m_2) ] +\frac{\lambda^2}{\alpha}{\mu^2} \\
B^j_{m_2\tilde m_2}&=&   \sqrt{(j+m_2+1)(j-m_2)(j+\tilde m_2+1)(j-\tilde m_2)}\\
H^j_{m_2\tilde m_2}&=& \sqrt{(j+m_2)(j-m_2+1)(j+\tilde m_2)(j-\tilde m_2+1)}.
\eeqa

 The use of the matrix basis for $\R^3_\lambda$ yields an interaction term which is diagonal (i.e a simple trace of product of matrices built from the coefficients of the fields expansion) whereas the kinetic term is not diagonal.
 
 In \cite{VW12} the scalar action could be written as an infinite sum of contributions, namely $S[\Phi]=\sum_{ j\in{{\mathbb{N} }\over{2}}}S^{(j)}[\Phi]$,
where  $S^{(j)}$  and describes a scalar action on the fuzzy sphere $\mathbb{S}^j$. Thus the problem essentially decoupled and could be solved in each $2j+1\times 2j+1$ subspace.  The kinetic term was an operator of Jacobi type (a three-diagonal matrix). It was therefore possible to invert it in terms of orthogonal polynomials, and obtain  the propagator $P^{j_1 j_2}_{mn;kl}$, defined by
\begin{equation}
\sum_{k,l=-j_2}^{j_2}\Delta^{j_1 j_2}_{mn;lk}P^{j_2 j_3}_{lk;rs}=\delta^{j_1 j_3}\delta_{ms}\delta_{nr},\ \sum_{m,n=-j_2}^{j_2}P^{j_1 j_2}_{rs;mn}\Delta^{j_2 j_3}_{mn;kl}=\delta^{j_1 j_3}\delta_{rl}\delta_{sk}\label{definvers}.
\end{equation}
The task was achieved by means of the dual Hahn polynomials, which are nothing but fuzzy spherical harmonics. They represent the basis which diagonalizes the kinetic term but not the interaction term.

Here the situation is much more complicated. The action does not decouple into an infinite block-diagonal form.  The kinetic term is not a Jacobi operator, since not only the indices $m,\tilde m$ vary in each subspace with fixed j, but j itself varies.

\section{Summary and Outlook}
We have proposed a new Laplacian on the algebra $\R^3_\lambda$ and computed the kinetic term for  a scalar action with quartic interaction. It is interesting to stress once again that the new Laplacian does not generalize trivially the Laplacian of fuzzy spheres, but contains a term which introduces, loosely speaking a dynamics between the spheres.  
There are however many interesting open problems yet to be addressed. 
\begin{itemize}
\item In order to perform quantum calculations, we need the propagator. This amounts to invert the kinetic action, which, differently from our previous experience, is not of Jacobi type, therefore it cannot be inverted with standard techniques involving orthogonal polynomials. 
 
    \item Besides this problem, which might be difficult to solve, but is essentially of technical nature, the issue of the commutative limit for our operator is for the moment unclear to us. In order to perform it correctly it is not enough to replace star products with commutative products and $\lambda=0$ everywhere, since the fields depend on $\lambda$ and the matrix basis is homogeneous in $x/\lambda$. A rescaling procedure should be  probably more appropriate.

\item Another interesting direction is the possibility of defining an outer derivation representing  the dilation, which is a star-derivation at the price of enlarging the noncommutative algebra with the parameter $\lambda$ itself. This was already considered in the literature \cite{GLRV06} and amounts not to fix the  noncommutative parameter in the spirit of \cite{Doplich1}. With the aid of this outer derivation it might be easier to construct a Laplacian which as a standard commutative limit, but many problems, such as the nature of the algebra and the integration measure need to be fixed.
    \end{itemize}

\noindent
{\bf{Acknowledgments}}: We thank  F. Lizzi, G. Marmo, P. Martinetti and J.-C. Wallet  for discussions and constructive comments. 


\begin{thebibliography}{50}
\bibitem{wiki} {http://en.wikipedia.org/wiki/Noncommutative$\_$quantum$\_$field$\_$theory}

\bibitem{Doplich1} S. Doplicher, K. Fredenhagen and J. E. Roberts, ''Space-time quantization induced by classical gravity ''Phys. Lett. B{\bf{331}} (1994) 39. \\
    S.~Doplicher, K.~Fredenhagen and J.~E.~Roberts,
  ``The Quantum structure of space-time at the Planck scale and quantum fields,''
  Commun.\ Math.\ Phys.\  {\bf 172} (1995) 187
  [hep-th/0303037].
\bibitem{Bronstein} M.~Bronstein,
  ``Quantum theory of weak gravitational fields,''
  Gen.\ Rel.\ Grav.\  {\bf 44}, 267 (2012).
 \bibitem{LQG}
  C.~Rovelli and L.~Smolin,
  ``Discreteness of area and volume in quantum gravity,''
  Nucl.\ Phys.\ B {\bf 442}, 593 (1995)
  [Erratum-ibid.\ B {\bf 456}, 753 (1995)]
  [gr-qc/9411005].

\bibitem{SW}
  N.~Seiberg and E.~Witten,
  ``String theory and noncommutative geometry,''
  JHEP {\bf 9909}, 032 (1999)
  [hep-th/9908142].
  \bibitem{lust}
  O.~Hohm, D.~Lust and B.~Zwiebach,
  ``The Spacetime of Double Field Theory: Review, Remarks, and Outlook,''
  Fortsch.\ Phys.\  {\bf 61}, 926 (2013)
  [arXiv:1309.2977 [hep-th]].
 
  
 
  \bibitem{Minwalla}
  S.~Minwalla, M.~Van Raamsdonk and N.~Seiberg,
``Noncommutative perturbative dynamics,''
  JHEP {\bf 0002} (2000) 020
  [arXiv:hep-th/9912072]; A.~Matusis, L.~Susskind and N.~Toumbas,
``The IR/UV connection in the non-commutative gauge theories,''
  JHEP {\bf 0012} (2000) 002
  [arXiv:hep-th/0002075]; I.~Chepelev and R.~Roiban,
``Renormalization of quantum field theories on noncommutative
  $\mathbb{R}^d$.  I: Scalars,'' JHEP {\bf 0005} (2000) 037
  [arXiv:hep-th/9911098].
  
  \bibitem{Grosse:2003aj}
  H.~Grosse and R.~Wulkenhaar, ''Power-counting theorem for non-local matrix models and renormalisation'',
  Commun.\ Math.\ Phys.\  {\bf 254} (2005) 91; H.~Grosse and R.~Wulkenhaar, ''Renormalisation of $\phi^4$-theory on noncommutative
  $\mathbb{R}^2$ in the matrix base'',
  JHEP {\bf 0312} (2003) 019; H.~Grosse and R.~Wulkenhaar, ''Renormalisation of $\phi^4$-theory on noncommutative
  $\mathbb{R}^4$ in the matrix base'',
  Commun.\ Math.\ Phys.\  {\bf 256} (2005) 305.

\bibitem{douglas} M.~R.~Douglas and N.~A.~Nekrasov,
  ``Noncommutative field theory,''
  Rev.\ Mod.\ Phys.\  {\bf 73}, 977 (2001)
  [hep-th/0106048].


\bibitem{szabo}
R. J. Szabo, ``Quantum field theory on noncommutative spaces,"
Phys.\ Rept.\  {\bf 378}, 207 (2003)
  [hep-th/0109162].

  
\bibitem{VW12} P.~Vitale and J.~-C.~Wallet,
  ``Noncommutative field theories on $R^3_\lambda$: Towards UV/IR mixing freedom,''
  JHEP {\bf 1304}, 115 (2013)
  [arXiv:1212.5131 [hep-th]].
 
\bibitem{GVW14}  A.~G\'er\'e, P.~Vitale and J.~-C.~Wallet,
  ``Quantum gauge theories on noncommutative 3-d space,''
  arXiv:1312.6145 [hep-th].
  \bibitem{Hammaa} A.~B.~Hammou, M.~Lagraa and M.~M.~Sheikh-Jabbari,
  ``Coherent state induced star product on R**3(lambda) and the fuzzy sphere,''
  {Phys.\ Rev.\ D}  {\bf 66}, 025025 (2002)
  [arXiv:hep-th/0110291].
\bibitem{selene}
J.~M.~Gracia-Bond\'ia, F.~Lizzi, G.~Marmo and P.~Vitale,
  ``Infinitely many star products to play with,''
  \emph{JHEP} {\bf 0204}, 026 (2002)
  [arXiv:hep-th/0112092];
      \bibitem{KS} P. Kustaanheimo and E. Stiefel, ''Perturbation Theory of
Kepler Motion Based on Spinor Regularization'',  J. Reine Angew. Math. {\bf 218} (1965) 204.
\bibitem{DMV05} A.~D'Avanzo, G.~Marmo and A.~Valentino,
  ``Reduction and unfolding for quantum systems: The Hydrogen atom,''
  Int.\ J.\ Geom.\ Meth.\ Mod.\ Phys.\  {\bf 2}, 1043 (2005)
  [math-ph/0504033].

\bibitem{Wick-Voros} A. Wick-Voros, ``Wentzel-Kramers-Brillouin method in the Bargmann representation'',
Phys. Rev. A {\bf 40} 6814 (1989).
\bibitem{discofuzzy}  F.~Lizzi, P.~Vitale and A.~Zampini, ``The Fuzzy disc,''
  JHEP {\bf 0308} (2003) 057 [hep-th/0306247].
   ``The Beat of a fuzzy drum: Fuzzy Bessel functions for the disc,''
  JHEP {\bf 0509} (2005) 080 [hep-th/0506008].
\bibitem{V-T} A.~Tanasa and P.~Vitale,
  ``Curing the UV/IR mixing for field theories with translation-invariant $\star$ products,''
  Phys.\ Rev.\ D {\bf 81}, 065008 (2010)
  [arXiv:0912.0200 [hep-th]].

\bibitem{pepejoe}
J.~M. Gracia-Bond{\'\i}a and J.~C. V{\'a}rilly, ``Algebras of distributions suitable for phase space
  quantum mechanics. {I},'' J.\ Math.\ Phys.\ {\bf 29} (1988) 869; ``Algebras of distributions suitable for phase-space quantum mechanics. {II}. Topologies on the
Moyal algebra,''
  J.\ Math.\ Phys.\  {\bf 29} (1988) 880.

\bibitem{GLRV06} J.~M.~Gracia-Bondia, F.~Lizzi, F.~R.~Ruiz and P.~Vitale,
  ``Noncommutative spacetime symmetries: Twist versus covariance,''
  Phys.\ Rev.\ D {\bf 74} (2006) 025014
   [Erratum-ibid.\ D {\bf 74} (2006) 029901]
  [hep-th/0604206].


\end{thebibliography}
\end{document}